\documentstyle[12pt]{article}

   \font\tenmsb=msbm10 scaled\magstep 1
   \font\sevenmsb=msbm7 scaled \magstep 1
   \font\faivemsb=msbm5 scaled \magstep 1
\newfam\msbfam
      \textfont\msbfam=\tenmsb
      \scriptfont\msbfam=\sevenmsb
      \scriptscriptfont\msbfam=\faivemsb
\def\Bbb#1{{\fam\msbfam #1}}
\font\tengothic=eufm10 scaled\magstep 1
\font\sevengothic=eufm7 scaled\magstep 1
\newfam\gothicfam
      \textfont\gothicfam=\tengothic
      \scriptfont\gothicfam=\sevengothic

\newcommand{\lbd}{\lambda}
\newcommand{\Om}{\Omega}
\newcommand{\om}{\omega}
\newcommand{\ra}{\rightarrow}
\newcommand{\prt}{\partial}
\newcommand{\be}{\begin{equation}}
\newcommand{\ee}{\end{equation}}
\newcommand{\dlt}{\delta}
\newcommand{\al}{\alpha}
\newcommand{\ep}{\varepsilon}
\newcommand{\dgr}{\dagger}

\begin{document}

\begin{center}
{\Large{\bf Possibility of Turbulent Crystals} \\ [5mm]

V.I. Yukalov$^1$ and E.P. Yukalova$^2$} \\ [2mm]

{\it
$^1$Bogolubov Laboratory of Theoretical Physics \\
Joint Institute for Nuclear Research \\
Dubna 141980, Russia \\ [2mm]

$^2$Laboratory of Informational Technologies \\
Joint Institute for Nuclear Research \\
Dubna 141980, Russia}

\end{center}

\vskip 3cm

\begin{abstract}

The possibility for the occurrence in crystals of a phenomenon, resembling 
turbulence, is discussed. This phenomenon, called {\it heterophase 
turbulence}, is manifested by the fluctuational appearance inside a 
crystalline sample of disordered regions randomly distributed in space. The 
averaged picture for such a turbulent solid is exemplified by an exactly 
solvable lattice-gas model. The origin of heterophase turbulence is 
connected with stochastic instability of quasi-isolated systems.

\end{abstract}

\newpage

\section{Typical Stages of Motion}

The term turbulence is usually applied to characterizing chaotic motion of 
fluids [1]. Therefore, it could sound strange to mention turbulence in 
connection with solids. Nevertheless, a similar phenomenon can occur in the 
latter, being related to {\it mesoscopic heterophase fluctuations} [2].

We shall keep in mind everywhere below the word turbulence in its standard 
sense [3] as characterizing the motion possessing the following main 
stochastic properties: (i) sensitive dependence on initial conditions, due 
to exponential divergence of initially close trajectories; (ii) chaotic 
space structure at fixed moments of time; (iii) existence of a strange 
attractor. Note that Ruelle [4] used the title of turbulent crystal with 
regard to a hypothetical stable amorphous solid. However, real amorphous 
solids are always only metastable.

Turbulence is a spatio-temporal phenomenon. In order to describe  spatial 
and temporal scales typical of heterophase turbulence, let us recall what 
are, in general, characteristic scales for a system of interacting particles.

The first evident characteristic length is an effective {\it interaction 
radius} $r_{int}$ or a scattering length $a_s\sim r_{int}$. If a 
characteristic particle velocity is $v $, then the {\it interaction time} is
$$
\tau_{int} \sim \; \frac{r_{int}}{v} \; .
$$
In the temporal interval $0<t<\tau_{int}$, called the {\it dynamic stage}, 
particles move, to a first glance, randomly.

Another important length is a {\it mean free path} $\lambda_{mfp}$ defining 
the {\it local-equilibrium time}
$$
\tau_{loc} \sim \; \frac{\lbd_{mfp}}{v} \; .
$$
And the temporal diapason $\tau_{int}<t<\tau_{loc}$ is the {\it kinetic 
stage}. At this stage, because of mutual interactions, interparticle 
correlations begin arising. For the density of particles $\rho\sim a^{-3}$, 
with $a$ being a mean interparticle distance, the mean free path is 
$\lbd\sim(\rho a_s^2)^{-1}$, that is, $\lbd\sim a^3/a_s^2$. Therefore,
$$
\tau_{loc} \sim \left ( \frac{a}{r_{int}}\right )^3 \tau_{int} \; .
$$
In condensed matter, $a\sim r_{int}$, because of which 
$\tau_{loc}\sim\tau_{int}$, and the kinetic stage being practically absent.

After the local equilibrium time $\tau_{loc}$, the state of local 
equilibrium is being evolved, and for $t>\tau_{loc}$, one has the {\it 
hydrodynamic stage}. But the latter consists of two qualitatively 
different parts. There exists a {\it coherence radius} $r_{coh}$ 
characterizing the effective size of mesoscopic heterophase fluctuations. 
The related {\it coherence time} is
$$
\tau_{coh} \sim \; \frac{r_{coh}}{v} \; .
$$
Heterophase fluctuations are mesoscopic because the coherence radius is 
between the mean interparticle distance $a$ and the linear size of the 
whole system $L$,
$$
a\ll r_{coh} \ll L \; .
$$
The coherence time $\tau_{coh}\gg \tau_{loc}$ defines an average lifetime 
of a mesoscopic fluctuation. The temporal interval $\tau_{loc}<t<\tau_{coh}$
corresponds to the {\it mesoscopic stage}, when heterophase fluctuations 
look as frozen, making the system nonuniform.

For times $t>\tau_{coh}$, heterophase fluctuations become dynamic. They 
arise randomly in time, being chaotically located in space. Their 
appearance signifies the existence of local instabilities leading to 
divergence of initially close trajectories. At each fixed moment of time, 
the spatial structure of the system is chaotic. Dynamics of a system with 
heterophase fluctuations corresponds to the evolution on a strange 
attractor [2]. These characteristics are the main features of a 
turbulent motion [3]. This is why the occurrence of heterophase 
fluctuations can be called {\it heterophase turbulence}. This phenomenon 
can be noticed only on the {\it macroscopic stage} during times 
$t>\tau_{coh}$.

Heterophase fluctuations are rather ubiquitous in nature [2]. They can be 
clearly observed by various experimental techniques. For instance, in 
many magnetic alloys, they have been studied by neutron scattering [5]. 
Mesoscopic heterophase fluctuations are common in liquid crystals [6], 
high-temperature superconductors [7--10], and many other materials [2]. 
Anomalous atomic diffusion in solids [11] and anomalous transport in some 
dynamic systems [12] can be explained by mesoscopic fluctuations. Such an 
anomalous diffusion is an analog of the turbulent diffusion [3]. Highly 
unusual properties of colossal magnetoresistive manganites are due to the 
coexisting fluctuating domains of semiconducting nonmagnetic states and 
metallic ferromagnetic states [13]. A good example of such a colossal 
magnetoresistive manganite is La$_{1-x}$Ca$_x$M$_n$O$_3$, with 
$x\approx 0.33$, where the fraction of the material in each of two 
well-defined states strongly depends on temperature and magnetic field, 
the inhomogeneity being characterized by a mesoscopic scale of about 100
$\AA$. Dynamical Jahn-Teller distortions around the antiferromagnetic to 
paramagnetic phase transition in UO$_2$ [14] and N$_p$O$_2$ [15] are also a 
kind of mesoscopic fluctuations. Such local fluctuations arise in many 
liquids, including water, as mesoscopic cooperative structures of 
different polymorphism [2,16--20], making a system dynamically 
heterogeneous. Seismic precursors related to local polymorphic 
transformations [21] can be viewed as mesoscopic heterophase phenomena. 
This concerns as well the appearance of droplets of quark-gluon plasma in 
a bulk hadronic medium and of hadronic bubbles in a bulk quark-gluon 
plasma, occurring around the quark-hadron transition [22,23]. Let us 
stress that heterophase fluctuations should not be confused with formally 
introduced microscopic clusters in the so-called equilibrium cluster 
approximations [24,25], but these fluctuations represent nonequilibrium 
mesoscopic germs with properties qualitatively differing from the 
surrounding matter.

\section{Averaged Picture of Turbulence}

There exists a statistical approach to describing turbulent motion in terms 
of an averaged picture [1,3]. Similarly to this, it is possible to develop 
an averaged picture of heterophase turbulence by averaging over heterophase 
fluctuations [2].

The appearance of turbulence in fluids happens when the Reynolds number 
surpasses a critical value. Analogously, heterophase turbulence arises only 
under a specific combination of interaction parameters and external 
variables, such as temperature or pressure. The Reynolds number for fluids 
is the ratio $vL/\nu$, in which $v$ is a characteristic velocity of motion, 
$L$ is a characteristic linear size, and $\nu$ is kinematic viscosity. That 
is, the Reynolds number is the ratio of factors favoring turbulence to 
those stabilizing laminar motion. A counterpart of the Reynolds number for 
heterophase turbulence could be the ratio of interaction parameters 
favouring a disordered state to those stabilizing an ordered state. 
Although the definition of such a typical ratio for heterophase turbulence 
is not always simple and, in addition, depends on thermodynamic variables.

In crystals, heterophase turbulence is manifested by the emergence of 
chaotically distributed regions of crystalline disorder [16,17]. Factors 
favoring the appearance of such regions of disorder could be a high 
concentration of defects and strong anharmonicity. Good candidates for this 
could be quantum crystals [26] and strongly anharmonic crystals [27,28]. 
The realistic description of such crystals involves rather complicated 
techniques. In order to avoid mathematical complications, we shall resort 
here to a simple lattice-gas model [29] serving as a cartoon of a crystal 
with many defects or of a solid with pores and cracks [30]. An advantage of 
employing this model is the possibility to suggest its rigorous analysis 
and to explicitly demonstrate that heterophase turbulence can correspond to 
a more stable state than a uniform homophase state.

Consider a system of $N$ particles in volume $V$. Particles are distributed 
on a crystalline lattice so that each lattice site can be occupied by not 
more than one particle. The number of lattice sites $N_s$ can be larger 
than that of particles, $N_s\gg N$; thus vacant sites can be present. 
Assume that the distribution of vacancies is not uniform over the system 
but that vacancies fluctuationally group in some parts of the sample, as a 
result of which the system possesses essentially different density of 
particles in different spatial regions, the latter being distributed 
chaotically. Such a picture is typical of heterophase turbulence, when a 
more dense phase and more dilute phase randomly fluctuate in space and time.

In an averaged description, the total number of particles can be presented 
as the sum $N=N_1+N_2$ of particles pertaining to the dense phase, $N_1$, 
and of those in the dilute phase, $N_2$. The whole volume of the system 
consists of two parts, $V=V_1+V_2$, those occupied by the dense and dilute 
phases, respectively. Each volume $V_\nu$, with $\nu=1,2$, does not form a 
connected part filled by the corresponding phase, but is composed of a 
large number of disconnected mesoscopic regions randomly located in space.

Since the number of sites, $N_s$, is larger than the total number of particles,
$N$, the density of sites is higher than that of particles,
\be
\rho_s >\rho \; , \qquad \rho_s \equiv \frac{N_s}{V} \; , \qquad
\rho\equiv \frac{N}{V}\; .
\ee
By definition, the dense phase is characterized by a higher density of 
particles that the dilute phase,
\be
\rho_1>\rho_2 \; , \qquad \rho_\nu \equiv \frac{N_\nu}{V_\nu} \qquad
(\nu=1,2) \; .
\ee
The geometric probabilities of the corresponding phases are
\be
w_\nu \equiv \frac{V_\nu}{V} \qquad 
( w_1 + w_2 =1 \; , \;\; 0\leq w_\nu \leq  1)\; .
\ee
The average density of particles can be written as
\be
\rho=w_1\rho_1 + w_2\rho_2 \; .
\ee
It is convenient to introduce the dimensionless densities
\be
n_\nu \equiv \frac{\rho_\nu}{\rho_s} \; \qquad n\equiv 
\frac{\rho}{\rho_s}\; .
\ee
Then Eq. (4) reads
\be
n= w_1n_1 + w_2 n_2 \; .
\ee

To an $i$-site of the lattice, pertaining to a $\nu$-phase, one ascribes 
the variable $e_{i\nu}$ taking the value $e_{i\nu}=1$ when the site is 
occupied by a particle and the value $e_{i\nu}=0$ if the site is empty. The 
phase densities can be presented through the averages
\be
\rho_\nu =\frac{1}{V} \; \sum_{i=1}^{N_s} <e_{i\nu}> \; , \qquad
n_\nu =\frac{1}{N_s}\; \sum_{i=1}^{N_s}<e_{i\nu}> \; ,
\ee
where $<\dots>$ implies a statistical averaging.

The procedure of averaging over heterophase fluctuations [2], for the 
considered case, yields the effective {\it heterophase Hamiltonian}
\be
\overline H = H_1\oplus H_2 \; ,
\ee
being a direct sum of the {\it phase-replica Hamiltonians}
\be
H_\nu = w_\nu\sum_{i=1}^{N_s} \left ( K_i - \mu\right ) e_{i\nu} +
\frac{1}{2}\; w_\nu^2 \; \sum_{i\neq j}^{N_s}\; 
\Phi_{ij} e_{i\nu}e_{j\nu} \; ,
\ee
in which $K_i$ is kinetic energy, $\mu$ is chemical potential, and 
$\Phi_{ij}$ is interaction potential. The phase probabilities $w_\nu$ are 
defined by minimizing the grand potential
\be
\Omega = - \Theta\; \ln {\rm Tr}\; e^{-\beta\overline H} \qquad
(\beta\Theta\equiv 1) \; ,
\ee
where $\Theta$ is average temperature in energy units. With the notation
\be
w\equiv w_1 \; , \qquad  w_2 =1 - w \; ,
\ee
the minimization equations are
\be
\frac{\prt\Om}{\prt w} = 0 \; , \qquad \frac{\prt^2\Om}{\prt w^2} > 0 \; .
\ee
The first of these equations leads to the equality
\be
< \frac{\prt\overline H}{\prt w}>\; = 0 \; .
\ee
The second of Eqs. (12) tells that the {\it heterophase susceptibility}
\be
\chi_\Theta \equiv \frac{1}{N_s} \; \frac{\prt^2\Om}{\prt w^2} > 0
\ee
is to be positive. Differentiating the thermodynamic potential (10) gives
\be
\chi_\Theta =\frac{1}{N_s}\left [ <\frac{\prt^2\overline H}{\prt w^2}> \; -
\beta <\left ( \frac{\prt\overline H}{\prt w}\right )^2> \right ] \; .
\ee
The inequality $\chi_\Theta > 0$ is a condition of {\it heterophase stability}.

In order to illustrate in an explicit way the thermodynamic properties of 
the considered system, without plunging to mathematical complications, let 
us choose the long-range interaction potential $\Phi_{ij}=\Phi_{ij}(N_s)$ 
depending on the number of lattice sites so that
\be
\lim_{N_s\ra\infty} \Phi_{ij}(N_s) = 0 \; , \qquad
\lim_{N_s\ra\infty} \left | \frac{1}{N_s} \sum_{i\neq j}^{N_s}
\Phi_{ij}(N_s)\right | < \infty\; .
\ee
This type of interaction potentials allows one to get, in the thermodynamic 
limit, asymptotically exact solutions [31], which is due to the 
asymptotically exact decoupling
\be 
<e_{i\nu}\; e_{j\nu}> \; = \delta_{ij}\; < e_{i\nu}>\; + \; (1-\delta_{ij})\;
<e_{i\nu}><e_{j\nu}> \; .
\ee

For the compactness of the following expressions, let us introduce the notation
\be
K\equiv \frac{1}{N_s} \; \sum_{i=1}^{N_s} \; K_i \; , \qquad
\Phi\equiv \frac{1}{N_s}\; \sum_{i\neq j}^{N_s} \; \Phi_{ij} \; ,
\ee
and let us define the dimensionless temperature and an effective chemical 
potential, respectively,
\be
T\equiv \frac{\Theta}{\Phi} \; ,\qquad \mu_* \equiv \frac{\mu-K}{\Phi} \; .
\ee
Also, denote the dimensionless densities (5) as
\be
n_1 \equiv a, \; \qquad n_2 \equiv b \qquad (a>b) \; .
\ee
Then for the grand potential (10), one obtains
$$
\frac{\Om}{\Phi N_s} =\frac{a}{2}(1-a)w^2 +\frac{b}{2}(1-b)(1-w)^2 -\;
\frac{1}{2}\;\mu_* -
$$
\be
- T\ln\left [ 4{\rm cosh}\left ( \frac{aw^2-\mu_*w}{2T}
\right ) {\rm cosh}\; \frac{b(1-w)^2-\mu_*(1-w)}{2T}\right ] \; .
\ee

The heterophase susceptibility (15) can be reduced to the form
\be
\chi_\Theta = (n_1^2 + n_2^2)\Phi = (a^2 + b^2)\Phi \; .
\ee
The condition of heterophase stability, $\chi_\Theta>0$, is nothing but the 
simple inequality
\be
\Phi >0 \; .
\ee
Heterophase turbulence is possible for positive interaction strength 
$\Phi$, which plays here the role of the Reynolds number.

The condition of heterophase stability (23) is a {\it necessary}, but not
sufficient, condition for the occurrence of heterophase turbulence. To show 
that the latter really develops, one has to check the existence of solutions 
for all thermodynamic characteristics. We have accomplished such an analysis 
by means of numerical calculations. Figure 1 demonstrates the temperature 
dependence of the partial densities (20) for different average densities 
$n$, and Fig. 2 presents the temperature dependence of the probability of 
the dense phase. Physically meaningful solutions require that the 
probability $w$ be in the interval $0\leq w\leq 1$. Hence, heterophase 
turbulence can actually arise not for all densities and temperatures. 
The solutions for the partial densities (20), defined by the equations
$$
a=\frac{1}{2}\; - \; \frac{1}{2}\;{\rm tanh}\left [ b(n-b)
\frac{2ab-n(a+b)}{2(a-b)^3T}\right ] \; , 
$$
\be
b=\frac{1}{2}\; - \; \frac{1}{2}\;{\rm tanh}\left [ a(a-n)
\frac{2ab-n(a+b)}{2(a-b)^3T}\right ] \; ,
\ee
have sense only when the dense-phase probability
\be
w=\frac{n-b}{a-b} \qquad (a> b)
\ee
satisfies the inequalities $0\leq w\leq 1$. The latter, together with the 
condition $a>b$, yield the requirements
\be
b<n< a \; , \qquad 0 < n <0.323 \; .
\ee

Figures 3 and 4 show the effective chemical potential
\be
\mu_* =\frac{a^2(n-b) + b^2(n-a)}{(a-b)^2} \; ,
\ee
and the dimensionless pressure
\be
P=-\; \frac{1}{\rho_s\Phi} \; \frac{\prt\Om}{\prt V}
\ee
as functions of temperature. The dimensionless specific heat
\be
C_V = -\; \frac{T}{N_s\Phi}\; \frac{\prt^2\Om}{\prt T^2}
\ee
and isothermal compressibility
\be
\kappa_T = -\;\frac{1}{V}\left (\frac{\prt P}{\prt V}\right )^{-1}
\ee
are shown in Figs. 5 and 6, respectively. Their positiveness demonstrates
the thermal and mechanical stability of the heterophase system. Finally, 
Fig. 7 demonstrates that the dimensionless grand potential 
$\om_w\equiv\Om/N_s\Phi$ for the heterophase matter is lower than that 
for the pure system, $\om_1$, when $w=1$. This means that the state with 
heterophase turbulence is globally more stable than the state having no 
heterophase fluctuations.

In this way, there exists a region of thermodynamic variables, e.g. 
temperature $\Theta$ and density $\rho$, where the lattice system displaying 
heterophase turbulence is absolutely stable as a whole, despite that the 
emerging turbulence assumes local instability. That is, local instability 
renders the system more stable globally. 

\section{Origins of Heterophase Turbulence}

In general, there could be three causes producing heterophase turbulence, 
extrinsic, intrinsic, and one intermediate between the first two, which 
may be called stochastic [2].

The {\it extrinsic} cause presupposses the action of sufficiently strong
external forces making the system nonequilibrium. These could be either
specially designed time-dependent perturbations or uncontrolled random
interactions with surrounding. This is similar to the way of preparing a 
turbulent fluid by strongly shaking the system [1], as a result of which the
fluid becomes locally turbulent, i.e. nonuniform, but on the average it is
uniform and isotropic, with zero average velocity. In the case of a solid, 
regions of disorder inside it could be formed by shock waves, irradiation
by particle beams, electromagnetic fields etc. The possibility of producing 
heterophase turbulence by means of strong external forces looks rather evident, 
hence not as interesting. A more intriguing question is whether heterophase
turbulence could emerge without strong external fields or not.

The {\it intrinsic} mechanism for the formation of heterophase fluctuations
is based on the Boltzmann hypothesis [32] that in the entirely equilibrium 
system strongly nonequilibrium local fluctuations can appear in a spontaneous
way. This mechanism of the appearance of nonequilibrium heterophase 
fluctuations from an equilibrium state was also advocated by Frenkel 
[33]. However, there is a slight inconsistence in this way of thinking. 
Really, first, one presupposses the existence of equilibrium states. Then 
one assumes that an equilibrium state can spontaneously pass to a 
nonequilibrium one. But the latter implies that equilibrium states are not 
absolutely stable, that is, they, actually, do not exist as stable 
states. Following this way of reasoning, it would be more logical to 
postulate from the very beginning that an absolutely stable state, 
to which an isolated system tends with time, is, generally, not an equilibrium
state but a quasiequilibrium heterophase state [2].

The {\it stochastic} source generating heterophase turbulence is intermediate 
between the above two in the following sense. It accepts that completely 
isolated systems do not exist in reality, but any given system is always 
subject to uncontrollable  random perturbations from the external surrounding. 
At the same time, one can prepare an almost isolated, or quasi-isolated, 
system for which external stochastic influence, if not completely 
eliminated, but is essentially reduced. Nevertheless, even infinitesimally 
small random perturbations can provoke noticeable chaoticity in a 
quasi-isolated system, with chaotic features governed by the system 
internal properties [2,34].

The fact that no real system can be ideally isolated has been repeatedly 
emphasized [35--39]. Moreover, it has been stressed [40,41] that the 
concept of an isolated system is self-contradictory by its own - To 
realize the isolation, one has to employ technical devices acting on the 
system, and to check that the latter is kept isolated, one needs to 
resort to measuring instruments perturbing the system.

In this section we show that even negligibly small stochastic perturbations 
can leed to the development of chaoticity in the system. The evolution 
equations are, generally, partial differential equations. Let a dynamical 
system be described by a set of $y(t)=[y_i(x,t)]$ of functions 
$y_i(x,t)$ enumerated by $i=1,2,\ldots,d$, with a continuous variable 
$x\in\Bbb{D}$ denoting a set of spatial variables pertaining to a domain 
$\Bbb{D}$, and $t\in\Bbb{R}_+$ being time. The set $y(t)$, called the
{\it dynamical state}, can be treated as a vector-column labelled by 
the discrete index $i$ and by the continuous multivariable $x$. The 
velocity field $v(y,t)=[v_i(x,y,t)]$ is also a column in the space of $i$ 
and $x$. Any system of evolution equations can be presented in the normal 
form
$$
\frac{d}{dt}\; y(t) = v(y,t)\; .
$$
The velocity field $v(y,t)$ may contain differential and integral operations. 
The appearance of chaos is related to the problem of stability of evolution 
equations.

\subsection{Autonomous systems}

Before passing to more complicated cases, let us recall just in a few lines  
how the problem of stability is treated for autonomous systems of 
nonlinear partial differential equations. This introductory consideration 
is useful as far as in the standard textbooks one usually studies only 
the stability of nonlinear ordinary differential equations. At the same 
time this example will show how to deal with the matrix notation [34] 
that will be constantly used in what follows.

A dynamical system is autonomous if the velocity field does not depend 
explicitly on time, $v(y,t)=v(y)$. Then the system evolution is given by 
a set of equations \be
\frac{d}{dt} \; y(t) = v(y) \; ,
\ee
presented in the matrix form.

One looks for a stationary solutions $y^*$ defined by the zero velocity field,
\be
v(y^*)=0 \; , \qquad y^*=[y_i(x)] \; .
\ee
For a small deviation
\be
\dlt y^*(t) = y(t) - y^* 
\ee
from the stationary solution $y^*$, one gets a linearized equation
\be
\frac{d}{dt} \; \dlt y^*(t) =\hat J^*\dlt y^*(t) \; ,
\ee
in which $\hat J^*=[J_{ij}^*(x,x')]$ is a contraction matrix with the elements
\be
J_{ij}^*(x,x') \equiv \frac{\dlt v_i(x,y^*)}{\dlt y_j^*(x')} \; .
\ee
Looking for the solutions of the eigenproblem
\be
\hat J^* f_n = J^*_n f_n \; ,
\ee
where $n$ is a multi-index, one finds the eigenvalues $J_n^*$ and the 
eigenfunctions $f_n=[f_{ni}(x)]$ of the contraction matrix. Assume that 
the eigenvalues form an  orthonormalized and complete basis, $\{ f_n\}$, 
so that 
$$
f_m^+f_n =\dlt_{mn}\; , \qquad \sum_n f_nf_n^+ =\hat 1 \; .
$$
Note that in the explicit presentation, the matrix eigenproblem (36) reads
$$
\sum_j \int J_{ij}^*(x,x')\; f_{nj}(x')\; dx' = J_n^* f_{ni}(x) \; .
$$
Expanding the deviations
\be
\dlt y^*(t) =\sum_n\dlt c_n(t) f_n
\ee
over the basis $\{ f_n\}$, one has the equation
\be
\frac{d}{dt}\; \dlt c_n(t) = J_n^*\dlt c_n(t)
\ee
for the coefficients $\dlt c_n(t)$. This equation immediately gives
\be
\dlt c_n(t) = \dlt c_n(t_0)\exp\left\{ J_n^*(t-t_0)\right\} \; .
\ee

The stability of the stationary solutions is characterized by the Lyapunov
exponents
\be
\lbd_n \equiv \lim_{t\ra\infty} \; \frac{1}{t}\;
\ln\left | \frac{\dlt c_n(t)}{\dlt c_n(t_0)}\right | = {\rm Re}\; J_n^*\; .
\ee
The set $\{\lbd_n\}$ is the Lyapunov spectrum. The functions $f_n$ are termed
normal modes. One tells that an $n$-mode is stable if $\lbd_n<0$, neutral when 
$\lbd_n=0$, and unstable for $\lbd_n>0$. The motion is chaotic if at least one 
mode is unstable.

\subsection{Stochastic systems}

Let us now turn to the problem of stability of stochastic differential 
equations. Assume that the system is subject to the action of stochastic 
fields $\xi(t)$. We continue using the matrix notation, where the dynamical 
state $y(\xi,t)=[y_i(x,\xi,t)]$ and the velocity field 
$v(y,\xi,t)=[v_i(x,y,\xi,t)]$ are the columns in $i$ and $x$. The 
evolution of a stochastic dynamical system is given by the matrix equation
\be
\frac{d}{dt}\; y(\xi,t) = v(y,\xi,t) \; .
\ee
Let the evolution equation (41) be complimented by the initial condition
\be
y(\xi,0) = y(0) \; ,
\ee
where $y(0)$ is a set of given functions $y_i(x)$. To solve a stochastic 
differential equation implies to find the solution
\be
y(t) \equiv \ll y(\xi,t)\gg 
\ee
averaged over the stochastic fields $\xi(t)$. We denote the stochastic 
averaging by the double angle brackets $\ll\ldots\gg$, reserving the single 
brackets $<\ldots>$ for statistical averaging, as in Sec. 2.

Note that stochastic differential equations [42] can be defined either in 
the sense  of Ito or in the sense of Stratonovich. We prefer to use the 
latter definition which involves simpler calculations, agreeing with 
the standard analysis, and which is better motivated physically [43]. 
Another possibility could be to present the stochastic fields
$\xi(t)$ as expansions, with random coefficients over smooth functions of 
time [44,45]. This method makes it possible to employ the standard 
analysis in differentiating and integrating over time. The final results 
of the expansion approach [44,45] coincide with those of the Stratonovich 
way.

The stability of the stochastic equation (41) cannot be analysed in the same 
manner as the stability of the autonomous equation (31), since the stationary 
solutions for the stochastic equation (41) are not defined. Another approach 
is necessary. Let us introduce the {\it local stability factor}
\be
\nu(t)\equiv \sup_{\dlt y(0)} \; \frac{|\dlt y(t)|}{|\dlt y(0)|}
\ee
characterizing the maximal deviation of the average trajectory at time $t$ 
after an infinitesimal variation of initial conditions $\dlt y(0)\ra 0$. The 
following classification for the local property of motion is evident:
$$
\nu(t) < 1 \; \qquad locally\; stable;
$$
$$
\nu(t) = 1 \; \qquad locally\; neutral;
$$
\be
\nu(t) > 1 \; \qquad locally\; unstable \; .
\ee
If the limit
$$
\nu(\infty) = \lim_{t\ra\infty}\nu(t)
$$
exists, then one can tell that the motion is asymptotically stable for 
$\nu(\infty)<1$, asymptotically neutral for $\nu(\infty)=1$, and 
asymptotically unstable for $\nu(\infty)>1$. The quantity
$$
\lbd\equiv \lim_{t\ra\infty} \; \frac{1}{t}\ln \nu(t)
$$
is the largest Lyapunov exponent.

The usage of the local stability factor (44) provides one with a much richer 
information on the motion than the largest Lyapunov exponent, since it is 
often important to study local in time stability but not only the 
asymptotic, as $t\ra\infty$, property of motion [34]. Many dynamical 
systems display a very complicated structure of their phase space, 
containing different singular zones, such as trapping islands, because of 
which the fine local properties of orbits play a leading role, while 
fairly rough characteristics such as the Lyapunov exponent are less 
important [46]. In addition, for stochastic differential equations, the 
divergence of trajectories is not necessarily exponential [47]. For example, 
if $\nu(t)\sim t^\beta$, then the motion can be either asymptotically stable 
or unstable depending on whether $\beta<0$ or $\beta>0$, while the Lyapunov 
exponent $\lbd=0$ for both these cases characterizes the motion as neutral, 
thus, loosing information about the motion.

To calculate the stability factor (44), we need to find 
$\dlt y(t)\equiv\ll\dlt y(\xi,t)\gg$. For this purpose, we introduce the 
{\it stochastic multiplier matrix} [34] as a matrix
$\hat M(\xi,t)=[M_{ij}(x,x',\xi,t)]$ with the elements
\be
M_{ij}(x,x',\xi,t) \equiv \frac{\dlt y_i(x,\xi,t)}{\dlt y_j(x',0)}
\ee
and the contraction matrix $\hat J(\xi,t)=[J_{ij}(x,x',\xi,t)]$ with the 
elements
\be
J_{ij}(x,x',\xi,t) \equiv \frac{\dlt v_i(x,y,\xi,t)}{\dlt y_j(x',\xi,t)}\; .
\ee
Accomplishing the variational differentiation of the evolution equation (41), 
we get the equation
\be
\frac{d}{dt}\; \hat M(\xi,t) =\hat J(\xi,t)\hat M(\xi,t)
\ee
for the multiplier matrix. When for the latter the eigenproblem
\be
\hat M(\xi,t)\varphi_n(t)=\mu_n(\xi,t)\varphi_n(t)
\ee
is defined, then the stability factor (44) becomes
\be
\nu(t) =\sup_n |\ll \mu_n(\xi,t)\gg | \; .
\ee
The eigenvalues $\mu_n(\xi,t)$ of the stochastic multiplier matrix 
$\hat M(\xi,t)$  can be found from Eqs. (48) and (49).

\subsection{Quasi-isolated systems}

As is discussed in the beginning of this section, no real system can be 
ideally isolated, although the influence of random external perturbations 
could, probably, be made very small. Nevertheless, even arbitrarily small 
random perturbations could provoke system instability. To analyze this 
problem, we need to consider the stability of a given dynamical system not
only with respect to the variation of initial conditions but also with 
respect to small stochastic perturbations.

Let the amplitude of stochastic fields be small, which can be presented by 
writing an explicit small factor $\al$ in front of $\xi(t)$. That is, 
instead of the evolution equation (36), we consider
\be
\frac{d}{dt}\; y(\al\xi,t) = v(y,\al\xi,t) \qquad (\al\ra 0) \; ,
\ee
where the complex parameter $\al$ is infinitesimally small, so that 
$\al\ra 0$ implies $\al_1\equiv{\rm Re}\al\ra 0$ and 
$\al_2\equiv{\rm Im}\al\ra 0$. When $\al\equiv 0$, no stochastic fields 
act on the system, and the evolution equation $$
\frac{d}{dt}\; y(0,t) = v(y,0,t)
$$
describes an isolated system. Switching on a small parameter models 
uncontrollable action of the surrounding. The parameter $\al$ is taken to 
be complex in order to simulate fluctuations of different physical 
quantities, for instance, of energy and attenuation, or of density and phase.

The stability of a stochastic system is characterized by the stability factor 
(50), which for Eq. (51) takes the form
\be
\nu(\ep,t) =\sup_n |\ll \mu_n(\al\xi,t)\gg|\; ,
\ee
where $\ep=\ep(\al_1,\al_2)$ is a function of $\al_1={\rm Re}\al$ and  
$\al_2={\rm Im}\al$, such that $\ep\ra 0$ as $\al\ra 0$. Depending on how 
$\al_1\ra 0$ and $\al_2\ra 0$, the parameter $\ep$ can behave either as 
$\ep\ra -0$ or $\ep\ra +0$, or it can even be that $\ep\equiv 0$. Of all 
these possibilities, one has to choose the case corresponding to the
maximal stability factor (52), as far as the latter, by definition, 
characterizes the maximal possible divergence of initially close trajectories. 
Therefore, everywhere in what follows, the limit $\ep\ra 0$ is defined as
\be
\lim_{\ep\ra 0} \equiv \sup_\ep\lim_{|\ep|\ra 0} \; .
\ee

To analyze the asymptotic stability one has to look at the temporal limit 
$t\ra\infty$. The result of taking the double limit of the stability 
factor (52) may depend on the order in which the limits $\ep\ra 0$ and 
$t\ra\infty$ are taken. We shall say [34] that the system is {\it 
stochastically stable} if the order of these limits is not important, 
that is, the limits commute with each other,
\be
[\lim_{t\ra\infty}\;, \lim_{\ep\ra 0}]\nu(\ep,t) = 0 \; .
\ee
Conversely, the system is termed {\it stochastically unstable}, when these 
limits do not commute:
\be
[\lim_{t\ra\infty}\;, \lim_{\ep\ra 0}]\nu(\ep,t) \neq 0 \; .
\ee
If a quasi-isolated system is stochastically unstable then, since any real 
system is never completely isolated but can be only quasi-isolated, chaos in 
the system can be induced by infinitesimally small random perturbations.

\section{Stochastic Density Matrices}

Now, employing the technique of the previous section, let us investigate if the 
evolution of particle density could be stochastically unstable. If so, this 
would mean that the density as a function of space and time could exhibit 
nonequilibrium fluctuations. Each of such fluctuations, after appearing, 
lives a finite lifetime, and then disappears. But, as time tends to 
infinity, the density does not tend to an equilibrium, say uniform or strictly 
periodic in space, state. As far as the system is stochastically unstable, 
nonequilibrium fluctuations arise again and again, chaotically emerging in 
time as well as in space. The state of matter with such chaotic fluctuations 
would be representing heterophase turbulence.

The density of particles
\be
\rho({\bf r},t) \equiv \lim_{{\bf r}'\ra{\bf r}}\; 
\rho_1({\bf r},{\bf r}',t)\; , 
\ee
as a function of the Cartesian vector ${\bf r}$ and time $t$, is a diagonal 
element of the single density matrix
\be
\rho_1({\bf r},{\bf r}',t) \equiv <\psi^\dgr({\bf r}',t)\psi({\bf r},t)>\; ,
\ee
in which the brackets $<\ldots>$ imply statistical averaging and 
$\psi({\bf r},t)$ is a field operator, being a vector-column in the space 
of internal degrees of freedom, such as spin, isospin, and so on. The evolution 
of the single density matrix (57) is coupled with that of the double density 
matrix
\be
\rho_2({\bf r}_1,{\bf r}_2,{\bf r}_1',{\bf r}_2',t) \equiv
<\psi^\dgr({\bf r}_2',t)\psi^\dgr({\bf r}_1',t)
\psi({\bf r}_1,t)\psi({\bf r}_2,t)>\; .
\ee
General properties of density matrices can be found in the book [48] (for 
some recent developments see Refs. [49,50]).

The evolution equation for the density matrix (57) can be presented [39] 
in the form
$$
i\; \frac{\prt}{\prt t}\; \rho_1({\bf r}_1,{\bf r}_2,t) = 
\left [ -\; \frac{1}{2m_0}
\left ( {\bf\nabla}_1^2 - {\bf\nabla}_2^2\right ) + 
U({\bf r}_1,t) - U({\bf r}_2,t)
\right ] \rho_1({\bf r}_1,{\bf r}_2,t) +
$$
\be
+ \int \left [ \Phi({\bf r}_1,{\bf r}_3) - \Phi({\bf r}_2,{\bf r}_3)\right ]
\rho_2({\bf r}_3,{\bf r}_1,{\bf r}_3,{\bf r}_2,t)\; d{\bf r}_3 \; ,
\ee
where we set the Planck constant $\hbar\equiv 1$, $m_0$ is particle mass, and
$U({\bf r},t)$ is an external potential, while 
$\Phi({\bf r},{\bf r}')=\Phi({\bf r}',{\bf r})$ is an interaction potential.

When ${\bf r}_2\ra{\bf r}_1$, then Eq. (59) reduces to the asymptotic form
\be
\frac{\prt}{\prt t}\; \rho_1({\bf r}_1,{\bf r}_2,t) = \frac{i}{2m_0}
\left ( {\bf\nabla}_1^2 - {\bf\nabla}_2^2\right ) 
\rho_1({\bf r}_1,{\bf r}_2,t)\; .
\ee
Taking account of the equality
$$
{\bf\nabla}_1^2 - {\bf\nabla}_2^2 = \left ( {\bf\nabla}_1 + {\bf\nabla}_2\right )
\left ( {\bf\nabla}_1 - {\bf\nabla}_2\right )
$$
and introducing [39] the density-of-currrent matrix
\be
{\bf j}({\bf r}_1,{\bf r}_2,t) \equiv -\; 
\frac{i}{2m_0} \left ( {\bf\nabla}_1 - {\bf\nabla}_2\right )
\rho_1({\bf r}_1,{\bf r}_2,t) \; ,
\ee
one can rewrite Eq. (60) as
\be
\frac{\prt}{\prt t}\; \rho_1({\bf r}_1,{\bf r}_2,t) = -
\left ( {\bf\nabla}_1 + {\bf\nabla}_2\right ){\bf j}({\bf r}_1,{\bf r}_2,t)\; .
\ee
The density of current is the diagonal element
\be
{\bf j}({\bf r},t) \equiv \lim_{{\bf r}'\ra{\bf r}}
{\bf j}({\bf r},{\bf r}',t)\; ,
\ee
which can also be represented as the statistical average
\be
{\bf j}({\bf r},t) = <\hat{\bf j}({\bf r},t)>
\ee
of the density-of-current operator
$$
\hat{\bf j}({\bf r},t) \equiv -\; \frac{i}{2m_0} \left [
\psi^\dgr{\bf\nabla}\psi - ({\bf\nabla}\psi^\dgr)\psi\right ] \; ,
$$
where, for brevity, we write $\psi\equiv\psi({\bf r},t)$. In the limit
${\bf r}_2={\bf r}_1$, Eq. (62) transforms to the standard continuity equation
\be
\frac{\prt}{\prt t}\; \rho_1({\bf r},t) - {\bf\nabla}\cdot{\bf j}({\bf r},t)
 = 0 \; .
\ee
Hence, we may say that Eq. (60) asymptotically, as ${\bf r}_2\ra{\bf r_1}$, 
equivalent to the continuity equaiton (65) defining the evolution of the 
particle density (56).

To check the stochastic stability of the evolution equation (60), one has to
incorporate there a small stochastic term. For the compactness of notation, let
us denote $x=\{{\bf r}_1,{\bf r}_2\}$. Then an asymptotically small stochastic
field is $\al\xi(x,t)$, with $\al\ra 0$. Also define
\be
y(x,\al\xi,t) \equiv \rho_1({\bf r}_1,{\bf r}_2,\al\xi,t) \; .
\ee
The stochastic equation, obtained from the evolution equation (60), reads
\be
\frac{\prt}{\prt t}\; y(x,\al\xi,t) = v(x,y,\al\xi,t) \; ,
\ee
where the velocity field is
\be
v(x,y,\al\xi,t) =\left [ \frac{i}{2m_0}\left ( {\bf\nabla}_1^2 - 
{\bf\nabla}_2^2\right ) +\al\xi(x,t)\right ] y(x,\al\xi,t) \; .
\ee
The stochastic term initiates here fluctuations of energy and of attenuation 
leading to the related fluctuations of the amplitude and phase of the 
density matrix.

Following the strategy of the previous section, we find the contraction 
matrix (47),
\be
J(x,x',\al\xi,t) = \left [ \frac{i}{2m_0}\left ( {\bf\nabla}_1^2 - 
{\bf\nabla}_2^2\right ) +\al\xi(x,t)\right ]\dlt(x-x') \; .
\ee
Since $\al\ra 0$, the eigenvalues of the matrix (69) can be found by means of
perturbation theory. In the zero approximation, the normalized eigenfunctions 
of the contraction matrix are
\be
\varphi_k(x) =\frac{1}{V}\; \exp\left\{ i({\bf k}_1\cdot{\bf r}_1 +
{\bf k}_2\cdot{\bf r}_2 )\right\} \; ,
\ee
with the notation $k=\{ {\bf k}_1,{\bf k}_2\}$. The eigenvalues of the 
contraction matrix, in the first order, are
\be
J_k(\al\xi,t) = \int \varphi_k^*(x) J(x,x',\al\xi,t)\varphi_k(x')\;
dx\; dx' \; .
\ee
This, with the functions (70), gives
\be
J_k(\al\xi,t) = -\; \frac{i}{2m_0}\left ( {\bf k}_1^2 -{\bf k}_2^2\right ) +
\al\xi_k(t) \; ,
\ee
where
\be
\xi_k(t) \equiv \int \varphi_k^*(x)\xi(x,t)\varphi_k(x)\; dx \; .
\ee
Note that if $\xi(x,t)=\xi(t)$ is a purely temporal perturbation, then 
Eqs. (70) and (72) give exact eigenfunctions and eigenvalues of the 
contraction matrix.

The eigenfunctions (70) are stationary, which means [34] that in this 
approximation the multiplier matrix (46) possesses the same eigenfunctions. 
The eigenvalues of the multiplier matrix (46) are given by the eigenproblem
\be
\int M(x,x',\al\xi,t)\varphi_k(x')\; dx' = \mu_k(\al\xi,t)\varphi_k(x) \; .
\ee
Taking into account the equation (48) for the multiplier matrix, one has
\be
\mu_k(\al\xi,t) =\exp\left\{ -\; \frac{i}{2m_0}\left ( k_1^2 - k_2^2\right ) t
+ \al\int_0^t \xi_k(t')\; dt'\right \} \; .
\ee
Assuming that $\xi(x,t)$ is a real Gaussian stochastic variable, centered at 
zero, for the stability factor (52), having here the form
$$
\nu(\ep,t) =\sup_k |\ll \mu_k(\al\xi,t)\gg | \; ,
$$
we obtain
\be
\nu(\ep,t) =\sup_k\exp\left\{ \ep \int_0^t \ll \xi_k(t')\xi_k(t'')\gg \;
dt'\; dt'' \right \} \; ,
\ee
with $\ep\equiv\frac{1}{2}(\al_1^2 -\al_2^2)$. Keeping in mind that 
$\ep\ra 0$ according to the limit (53), the factor (76) can be rewritten as
\be
\nu(\ep,t) =\exp\{ |\ep|\Gamma(t) t\} \; ,
\ee
where
\be
\Gamma(t) \equiv \sup_k\; \frac{1}{t}\left | \int_0^t
\ll \xi_k(t')\xi_k(t'')\gg \; dt'\; dt''\right | \; .
\ee

It is evident that
\be
\lim_{t\ra\infty}\lim_{\ep\ra 0}\nu(\ep,t) = 1 \; .
\ee
However, under fixed $|\ep|>0$, it may happen that the limit of $\nu(\ep,t)$ 
as $t\ra\infty$, does not exist, which could occur e.g. if $\Gamma(t)$ 
oscillates. Or it may be that $\Gamma(t)t\ra\infty$, as $t\ra\infty$, which 
results in $\nu(\ep,t)\ra\infty$. The latter behaviour can be illustrated by 
assuming that $\xi(x,t)$ describes the temporal white noise with the 
correlation function $$
\ll \xi(x,t)\xi(x',t')\gg = \gamma\dlt(t-t') \; ,
$$
where $\gamma>0$. Then $\Gamma(t)=\gamma$, and the factor (77) becomes
$$
\nu(\ep,t) =\exp(|\ep|\gamma t) \; .
$$
Another possibility could be to consider the uniform noise with the correlator
$$
\ll \xi(x,t)\xi(x',t')\gg \; = \gamma^2 \; .
$$
Then $\Gamma(t)=\gamma^2t$, and the factor (77) reduces to
$$
\nu(\ep,t) =\exp\left ( |\ep|\gamma^2 t^2 \right ) \; .
$$
In any of such cases,
\be
\lim_{\ep\ra 0}\lim_{t\ra\infty} \nu(\ep,t) = \infty \; ,
\ee
which demonstrates stochastic instability of the system, in agreement with 
definition (55).

\section{Conclusion}

The considered simple cases show that there always exist such infinitessimaly
small random perturbations which make the system stochastically unstable. 
In general, for stochastic instability, it is sufficient that
$$
\lim_{t\ra\infty} \Gamma(t)t = \infty\; .
$$
There exists a wide class of noises, for instance, ranging between the white 
noise and the uniform noise, which could provide the validity of this limit.

Instability of an evolution equation implies that there are no stationary 
solutions for this equation. A chaotic dynamical system can only possess 
a chaotic attractor. The above arguments show that the evolution of the 
particle density (56) can be stochastically unstable. Hence, the
density cannot tend, with time, to a stationary limit describing a 
uniform density for liquids or gases and a periodic density for crystals. 
But the density $\rho({\bf r},t)$ will always depend on time, occasionally 
displaying random spatio-temporal fluctuations of nonequilibrium 
nature. To be noticeable as nonequilibrium, these fluctuations have to be at 
least mesoscopic. Such nonequilibrium mesoscopic fluctuations make the 
system heterophase. The state of matter  with heterophase fluctuations 
represents heterophase turbulence.

In this sense, there should exist {\it turbulent crystals}, whose density 
displays chaotic spatio-temporal  fluctuations corresponding to mesoscopic 
regions of disorder. A turbulent crystal, being observed during the times 
shorter or of the order of the characteristic lifetime of heterophase 
fluctuations, $\tau_{coh}$, would be seen as a matter, in which
the spatial regions having approximately a crystalline structure with a type 
of lattice symmetry are randomly intermixed with disordered regions 
having no symmetry. The turbulent crystal, although being turbulent, is 
anyway a crystal because, being observed during the times much longer 
than the heterophase coherence time $\tau_{coh}$, possesses on the average 
a periodic density, with a well-defined symmetry of crystalline lattice. The 
corresponding density, averaged over such long times $t\gg\tau_{coh}$, can 
be written as 
$$
\overline\rho({\bf r}) \equiv \lim_{t\ra\infty} \frac{1}{t} \int_0^t 
\rho({\bf r},t')\; dt' \; .
$$
Assuming heterophase quasiergodicity [2], the average density writes
$$
\overline\rho({\bf r}) = w_1\rho_1({\bf r}) + w_2\rho_2 \; ,
$$
where $w_1$ and $w_2$ are statistical weights of the crystalline phase and of 
the disordered phase, respectively. The crystalline density $\rho_1({\bf r})$ 
is periodic, while that of the disordered phase is uniform. For the 
densities averaged over space, 
$$
\rho\equiv \int \overline\rho({\bf r})\; d{\bf r} \; , \qquad
\rho_\nu \equiv \int \rho_\nu({\bf r})\; d{\bf r} \; ,
$$
we return to the equality (4).

The possibility for the existence of turbulent crystals does not require 
that all crystals be necessarily turbulent. Whether they are such or not 
depends on the internal properties of the system as well as on external 
conditions, as is demonstrated in the model picture of section 2. But this 
possibility to become turbulent must be taken into account. Heterophase
turbulence may be absent at some particular conditions, but may emerge 
when changing the latter. For example, it may not exist at low temperatures, 
but can arise near the temperature of melting [16,17], where the very 
existence of the heterophase turbulence is necessary for understanding 
the physics of melting and crystallization.

\newpage

\begin{center}

{\bf Figure Captions}

\end{center}

{\bf Fig. 1}. The partial densities of the dense phase (solid curve) and of 
the dilute phase (dashed curve), for different average densities $n$, as 
functions of temperature.

\vskip 5mm

{\bf Fig. 2}. The probability of the dense phase versus temperature for 
different $n$.

\vskip 5mm

{\bf Fig. 3}. The effective chemical potential vs. temperature.

\vskip 5mm

{\bf Fig. 4}. The dimensionless pressure as a function of temperature.

\vskip 5mm

{\bf Fig. 5}. The dimensionless specific heat vs. temperature.

\vskip 5mm

{\bf Fig. 6}. The isothermal compressibility vs. temperature.

\vskip 5mm

{\bf Fig. 7}. The dimensionless grand potential for the heterophase matter, 
$\omega_w$, and for the pure dense system, $\omega_1$, vs. temperature.

\end{document}